\documentclass[12pt]{article}
  \usepackage{amsfonts}
  \usepackage{amsmath}
\usepackage{amssymb}
\usepackage{amscd}
\usepackage{graphicx}
\begin{document}

~~
\bigskip
\bigskip
\begin{center}
{\Large {\bf{{{Chaos synchronization of canonically and Lie-algebraically deformed Henon-Heiles systems by
active control}}}}}
\end{center}
\bigskip
\bigskip
\bigskip
\begin{center}
{{\large ${\rm {Marcin\;Daszkiewicz}}$}}
\end{center}
\bigskip
\begin{center}
\bigskip

{ ${\rm{Institute\; of\; Theoretical\; Physics}}$}

{ ${\rm{ University\; of\; Wroclaw\; pl.\; Maxa\; Borna\; 9,\;
50-206\; Wroclaw,\; Poland}}$}

{ ${\rm{ e-mail:\; marcin@ift.uni.wroc.pl}}$}

\end{center}
\bigskip
\bigskip
\bigskip
\bigskip
\bigskip
\bigskip
\bigskip
\bigskip
\bigskip
\begin{abstract}
Recently, there has been provided two chaotic models based on the twist-deforma-tion of classical Henon-Heiles system. First of them has been constructed on the well-known, canonical space-time noncommutativity, while the second one on the Lie-algebraically type of quantum space, with two spatial directions commuting to classical time. In this article, we find the direct link between mentioned above systems, by synchronization both of them in the framework of active control method. Particularly, we derive at the canonical phase-space level the corresponding active controllers as well as we perform (as an example) the numerical synchronization of analyzed models.
\end{abstract}
\bigskip
\bigskip
\bigskip
\bigskip
\eject

\section{{{Introduction}}}

Since Edward Lorenz proposed his widely-known ''model of weather'', there appeared a lot of papers dealing with so-called chaotic
models, whose dynamics is described by strongly sensitive with
respect initial conditions, nonlinear differential equations. The most popular of them
are: Lorenz system \cite{[1]}, Roessler system \cite{[2]}, Rayleigh-Benard system \cite{[3]}, Henon-Heiles
system \cite{[4]}, jerk equation \cite{[5]}, Duffing equation \cite{[6]}, Lotka-Volter system \cite{[7]}, Liu system
\cite{[8]}, Chen system \cite{[9]} and Sprott system \cite{[10]}. A lot of them have been applied in various
fields of industrial and scientific divisions, such as, for example: Physics, Chemistry,
Biology, Microbiology, Economics, Electronics, Engineering, Computer Science, Secure
Communications, Image Processing and Robotics.

The one of the most interesting among the above models seems to be so-called Henon-Heiles system, which has been provided in
pure astrophysical context. It concerns the problem of nonlinear motion of a star around of a galactic center, where the motion is
restricted to a plane. It is defined by the following Hamiltonian function
\begin{eqnarray}
H(p,x) \;=\; \frac{1}{2}\sum_{i=1}^2\; \left(p_i^2 + x_i^2\right) + x_1^2x_2 - \frac{1}{3}x_2^3\;,\label{ham}
\end{eqnarray}
which in cartesian coordinates $x_1$ and $x_2$ describes the set of two nonlinearly coupled harmonic oscillators.
In polar coordinates $r$ and $\theta$ it corresponds to the particle moving in noncentral potential of the form
\begin{eqnarray}
V(r,\varphi) \;=\; \frac{r^2}{2} + \frac{r^3}{3}\sin\left(3\varphi\right)\;,\label{pot}
\end{eqnarray}
with $x_1 = r\cos\varphi$ and $x_2 = r\sin\varphi$. The above model has been inspired by observations indicating, that star moving
in a weakly  perturbated central potential should has apart of  total energy $E_{\rm{tot}}$ constant in time, also the second
conserved physical quantity $I$\footnote{The quantity $I$ plays the role of additional constant of motion, which leads to the regular trajectories of particle.}. It has been demonstrated with use of so-called
Poincare section method, that such a situation appears in the case of Henon-Heiles system only for the values of control parameter $E_{\rm{tot}}$ below
the threshold $E_{\rm th} = 1/6$. For higher energies the trajectories in phase space become chaotic and the quantity $I$ does not exist (see e.g. \cite{tabor}, \cite{genhamhh}).

Recently, there has been proposed in articles \cite{d1} and \cite{d2} two noncommutative counterparts of the above mentioned Henon-Heiles system. They have been defined respectively on the following
canonically as well as Lie-algebraically deformed Galilei space-times \cite{oeckl}-\cite{dasz1}\footnote{The canonically and Lie-algebraically noncommutative space-times have been defined as the quantum
representation spaces, so-called Hopf modules (see e.g. \cite{oeckl}, \cite{chi}), for the twist-deformed quantum Galilei Hopf algebras $\,{\mathcal U}_\theta({ \cal G})$ and $\,{\mathcal U}_\kappa({ \cal G})$ respectively.},\footnote{It should be noted that in accordance with the
Hopf-algebraic classification  of all deformations of relativistic
and nonrelativistic symmetries (see references \cite{class1}, \cite{class2}),
apart of canonical \cite{oeckl}-\cite{dasz1} space-time noncommutativity, there also exist  Lie-algebraic \cite{dasz1}-\cite{lie1} and quadratic \cite{dasz1}, \cite{lie1}-\cite{paolo} type of quantum spaces.}
\begin{equation}
[\;t,\hat{x}_{i}\;] = 0\;\;\;,\;\;\; [\;\hat{x}_{i},\hat{x}_{j}\;] = 
i\theta_{ij}
\;,\label{nhspace}
\end{equation}
and
\begin{equation}
[\;t,\hat{x}_{i}\;] = 0\;\;\;,\;\;\; [\;\hat{x}_{i},\hat{x}_{j}\;] =
\frac{i}{\kappa}t \epsilon_{ij}
\;,\label{lie}
\end{equation}
with constant deformation parameters $\theta_{ij}=-\theta_{ji}$ and $\kappa$. Particularly, there has been provided the Hamiltonian functions of the models as well as the corresponding canonical equations of motion. Besides, it has been demonstrated that for proper values of deformation parameters $\theta$ and $\kappa$, and for proper values of control parameters, there appears (much more intensively) chaos in both systems. Consequently, in such a way, it has been shown the impact of the above noncommutative space-times on the basic dynamical properties of this important classical chaotic model. It should be noted, that such deformed constructions are inspired by investigations, dealing with noncommutative classical and quantum  mechanics (see e.g.
\cite{mech}-\cite{qmgnat}) as well as with field theoretical systems (see e.g. \cite{prefield}-\cite{fiorewess}), in which  the quantum
space-time is not classical. Such  models follow (particularly) from  formal  arguments based mainly  on
Quantum Gravity \cite{2}, \cite{2a} and String Theory
\cite{recent}, \cite{string1}, indicating that space-time at Planck
scale  becomes noncommutative.

One of the most important problem of the chaos theory concerns so-called chaos synchronization
phenomena. Since Pecora and Caroll \cite{[11]} introduced a method to synchronize
two identical chaotic systems, the chaos synchronization has received increasing attention
due to great potential applications in many scientific discipline. Generally, there are
known several methods of chaos synchronization, such as: OGY method \cite{[12]}, active control
method \cite{[13]}, \cite{[14]}, adaptive control method \cite{[17]}, \cite{[18]}, backstepping method \cite{[22]}, \cite{[23]},
sampled-data feedback synchronization method \cite{[24]}, time-delay feedback method \cite{[25]} and
sliding mode control method \cite{[26]}, \cite{[27]}. The mentioned methods have been applied to
the synchronization of many identical as well as different chaotic models, such as, for example, Sprott, Lorenz and Roessler
systems respectively \cite{[32a]}, \cite{[34a]}.

In this article we synchronize by active control scheme canonically deformed Henon-Heiles (master) system \cite{d1} with its Lie-algebraically noncommutative (slave) partner
\cite{d2}. In this aim we establish the proper so-called active controllers with use of the Lyapunov
stabilization theory \cite{[30]}. Additionally, we illustrate the obtained results by numerical calculations performed for particular values of deformation parameters
$\theta_{ij}$ and $\kappa$.

The paper is organized as follows. In second Section we recall chaotic canonically and Lie-algebraically deformed Henon-Heiles models proposed in
articles \cite{d1} and \cite{d2} respectively. In Section 3 we remaind the basic concepts of active synchronization method, while in
fourth Section we find
the active controllers which synchronize both noncommutative systems. The conclusions and
final remarks are discussed in the last Section.

\section{The noncommutative Henon-Heiles models}

In this Section we very shortly remaind the basic facts concerning two chaotic Henon-Heiles models defined on noncommutative Galilei space-times
(\ref{nhspace}) and (\ref{lie}) respectively. As  it was mentioned in Introduction, first of them has been provided in paper \cite{d1} while the second one in article \cite{d2}.

\subsection{Classical Henon-Heiles system on canonically deformed space-time}

In accordance with \cite{d1}, the dynamics of the model is given by the following Hamiltonian function
\begin{eqnarray}
H(\hat{p},\hat{x}) \;=\; \frac{1}{2}\sum_{i=1}^2\; \left(\hat{p}_i^2 + \hat{x}_i^2\right) + \hat{x}_1^2\hat{x}_2 - \frac{1}{3}\hat{x}_2^3\;,\label{nonham}
\end{eqnarray}
defined on the canonically deformed phase space of the form\footnote{The correspondence relations are $\{\;\cdot,\cdot\;\} = \frac{1}{i}\left[\;\cdot,\cdot\;\right]$.}
\begin{eqnarray}
&&\{\;\hat{ x}_{1},\hat{ x}_{2}\;\} = 2\theta\;\;\;,\;\;\;\{\;\hat{ p}_{1},\hat{ p}_{2}\;\} = \{\;\hat{ x}_{i},\hat{ p}_{j}\;\} =
0\;, \label{rel2}
\end{eqnarray}
with constant parameter ${\theta}= \theta_{12} = -\theta_{21}$. In terms of commutative canonical variables $({ x}_i, { p}_i)$ the Hamiltonian looks as follows
\begin{eqnarray}
H({p},{x}) &=&
\frac{1}{2M(\theta)}\left({{{p}}_1^2}+{{{p}}_2^2} \right)  +
\frac{1}{2}M(\theta)\Omega^2({\theta})\left({{{x}}_1^2}+{{{x}}_2^2} \right)
- S(\theta)L\;+ \label{2dh1}\\
&+& \left(x_1 - {{\theta}p_2}\right)^2\cdot \left(x_2 + {{ \theta}p_1}\right) -
\frac{1}{3}\left(x_2 + {{\theta}p_1}\right)^3\;,\nonumber
\end{eqnarray}
where
\begin{eqnarray}
&&L = x_1p_2 - x_2p_1\;, \\[5pt]
&&1/M({\theta}) = 1 +{\theta}^2 \;,\label{fazi1}\\[5pt]
&&\Omega({\theta}) = \sqrt{\left(1
+{\theta}^2 \right)}\;,
\end{eqnarray}
and
\begin{eqnarray}
S({\theta})={\theta}\label{fazi2}\;.
\end{eqnarray}
Due to the form of the above energy function, the symbols $M({\theta})$ and $\Omega({\theta})$ denote the new, deformed mass and frequency of particle, respectively. Obviously, quantity $L$ plays the role of the angular momentum vector, while $S(\theta)$ can be interpreted as the present in third term of the Hamiltonian, the new $\theta$-dependent coefficient. It should be also noted, that two last, nonlinear members of formula (\ref{2dh1}) remain responsible for chaotic behaviour of the system, while the corresponding to $H({p},{x})$ canonical equations
of motion are given by
\begin{eqnarray}
\dot{x}_1 &=& \left[{1}/{M({\theta})}\right]p_1 + S({\theta})x_2 + \nonumber\\
&~&~~~~~~~~~~~~~~~~~+\;\left[\left(x_1- {\theta}p_2\right)^2 - \left(x_2+ {\theta}p_1\right)^2\right]{\theta}\;,\\[10pt]
\dot{x}_2 &=& \left[{1}/{M({\theta})}\right]p_2 - S({\theta})x_1 - 2(x_2+\theta p_1)\left(x_1- {\theta}p_2\right){\theta}\;,\\[10pt]
\dot{p}_1 &=& -{M({\theta})}\Omega^2({\theta})x_1 + S({\theta})p_2 -
2\left(x_2+ {\theta}p_1\right)(x_1-\theta p_2)\;,\\[10pt]
\dot{p}_2 &=& -{M({\theta})}\Omega^2({\theta})x_2 - S({\theta})p_1 + \;\nonumber\\
&~&~~~~~~~~~~~~~~~~~~~~~~~~-\;\left(x_1- {\theta}p_2\right)^2 +
\left(x_2+ {\theta}p_1\right)^2\;.
\end{eqnarray}
Of course, for deformation parameter $\theta$ approaching zero the above system
becomes classical.

\subsection{Classical Henon-Heiles system on Lie-algebraically deformed space-time}

The model is defined by the Hamiltonian function (\ref{nonham}) given on the following Lie-algebraically deformed phase space
\begin{eqnarray}
\{\;\hat{ x}_{1},\hat{ x}_{2}\;\} = \frac{2t}{\kappa}\;\;\;,\;\;\;
\{\;\hat{ p}_{i},\hat{ p}_{j}\;\} = 0\;\;\;,\;\;\; \{\;\hat{ x}_{i},\hat{ p}_{j}\;\} = \delta_{ij}\;,\label{time1}
\end{eqnarray}
with constant, mass-like parameter ${\kappa}$\footnote{One can check that $[\kappa] = \rm{kg}$.}. In terms of commutative variables the above Hamiltonian takes the form
\begin{eqnarray}
H({p},{x},t) &=&
\frac{1}{2M\left(\frac{t}{\kappa}\right)}\left({{{p}}_1^2}+{{{p}}_2^2} \right)  +
\frac{1}{2}M\left(\frac{t}{\kappa}\right)\Omega^2\left(\frac{t}{\kappa}\right)\left({{{x}}_1^2}+{{{x}}_2^2} \right)
- S\left(\frac{t}{\kappa}\right)L\;+ \nonumber\\
&+& \left(x_1 - {\frac{t}{\kappa}p_2}\right)^2 \left(x_2 + {\frac{t}{\kappa}p_1}\right)
- \frac{1}{3}\left(x_2 + {\frac{t}{\kappa}p_1}\right)^3\;,\label{time2}
\end{eqnarray}
where
\begin{eqnarray}
&&L = x_1p_2 - x_2p_1\;, \\[5pt]
&&\frac{1}{M\left(\frac{t}{\kappa}\right)} = 1 +\left(\frac{t}{\kappa}\right)^2 \label{goblin1}\;,\\[5pt]
&&\Omega\left(\frac{t}{\kappa}\right) = \sqrt{\left(1
+\left(\frac{t}{\kappa}\right)^2 \right)}\;,
\end{eqnarray}
and
\begin{eqnarray}
S\left(\frac{t}{\kappa}\right)=\frac{t}{\kappa}\label{goblin2}\;.
\end{eqnarray}
It is worth to notice, that due to the similar form of energy functions (\ref{2dh1}) and (\ref{time2}), the all coefficients ${M\left(\frac{t}{\kappa}\right)}$, ${\Omega\left(\frac{t}{\kappa}\right)}$ as well as ${S\left(\frac{t}{\kappa}\right)}$ can be interpreted in the same manner as their $\theta$-deformed counterparts (\ref{fazi1})-(\ref{fazi2}). However, contrary to the pervious case, the Lie-algebraically modified quantities (\ref{goblin1})-(\ref{goblin2}) are time-dependent, and the corresponding canonical equations of motion look as follows
\begin{eqnarray}
\dot{x}_1 &=& {1}/{M\left(\frac{t}{\kappa}\right)}p_1 + S\left(\frac{t}{\kappa}\right)x_2 +\nonumber\\
&~&~~~~~~~~~~~~~~~~~+\;
\left[\left(x_1- \frac{t}{\kappa}p_2\right)^2 - \left(x_2+ \frac{t}{\kappa}p_1\right)^2\right]\frac{t}{\kappa}\;\\[5pt]
\dot{x}_2 &=& {1}/{M\left(\frac{t}{\kappa}\right)}p_2 - S\left(\frac{t}{\kappa}\right)x_1 - 2\left[x_2+ \frac{t}{\kappa}p_1\right]
\left[x_1- \frac{t}{\kappa}p_2\right]\frac{t}{\kappa}\;\\[5pt]
\dot{p}_1 &=& -{M\left(\frac{t}{\kappa}\right)}\Omega^2\left(\frac{t}{\kappa}\right)x_1 + S\left(\frac{t}{\kappa}\right)p_2 -
2\left[x_2+ \frac{t}{\kappa}p_1\right]\left[x_1- \frac{t}{\kappa}p_2\right]\;\\[5pt]
\dot{p}_2 &=& -{M\left(\frac{t}{\kappa}\right)}\Omega^2\left(\frac{t}{\kappa}\right)x_2 - S\left(\frac{t}{\kappa}\right)p_1 +\;\nonumber\\[5pt]
 &~&~~~~~~~~~~~~~~~~~~~~~~~~~-\;
 \left[x_1- \frac{t}{\kappa}p_2\right]^2 +
\left[x_2+ \frac{t}{\kappa}p_1\right]^2\;.
\end{eqnarray}
Obviously, for deformation parameter ${\kappa}$ running to infinity the above model becomes commutative.

\section{Chaos synchronization by active control - general prescription}

In this Section we remaind the general scheme of chaos synchronization of two systems
by so-called active control procedure \cite{[13]}, \cite{[14]}.
Let us start with the following master model\footnote{$\frac{do}{dt} = \dot{o}$.}
\begin{eqnarray}
\dot{x} = Ax + F(x) \;, \label{eq1}
\end{eqnarray}
where $x = [\;x_1, x_2,\dots ,x_n\;]$ is the state of the system, A denotes the $n \times n$ matrix of
the system parameters and $F(x)$ plays the role of the nonlinear part of the differential
equation (\ref{eq1}). The slave model dynamics is described by
\begin{eqnarray}
\dot{y} = By + G(y) + u \;,\label{eq2}
\end{eqnarray}
with $y = [\;y_1, y_2,\dots ,y_n\;]$ being the state of the system, $B$ denoting the $n$-dimensional
quadratic matrix of the system, $G(y)$ playing the role of nonlinearity of the equation (\ref{eq2})
and $u = [\;u_1, u_2,\dots ,u_n\;]$ being the active controller of the slave model. Besides, it should
be mentioned that for matrices $A = B$ and functions $F = G$ the states $x$ and $y$ describe
two identical chaotic systems. In the case $A \neq B$ or $F \neq G$ they correspond to the two
different chaotic models.

Let us now provide the following synchronization error vector
\begin{eqnarray}
e = y - x \;, \label{eq3}
\end{eqnarray}
which in accordance with  (\ref{eq1}) and  (\ref{eq2}) obeys
\begin{eqnarray}
\dot{e} = By - Ax + G(y) - F(x) + u\;. \label{eq4}
\end{eqnarray}

In active control method we try to find such a controller $u$, which synchronizes the state
of the master system (\ref{eq1}) with the state of the slave system (\ref{eq2}) for any initial condition
$x_0 = x(0)$ and $y_0 = y(0)$. In other words, we design a controller $u$ in such a way that for
system (\ref{eq4}) we have
\begin{eqnarray}
\lim_{t \to \infty}||{e}(t)|| =0\;, \label{eq5}
\end{eqnarray}
for all initial conditions $e_0 = e(0)$. In order to establish the synchronization (\ref{eq4}) we use
the Lyapunov stabilization theory \cite{[30]}. It means, that if we take as a candidate Lyapunov
function of the form
\begin{eqnarray}
V(e) = e^{T}Pe \;, \label{eq6}
\end{eqnarray}
with $P$ being a positive $n \times n$ matrix, then we wish to find the active controller $u$ so that
\begin{eqnarray}
\dot{V}(e) = -e^{T}Qe \;, \label{eq7}
\end{eqnarray}
where Q is a positive definite $n \times n$ matrix as well. Then the systems (\ref{eq1}) and (\ref{eq2}) remain
synchronized.

\section{Chaos synchronization of the models}

The described in pervious Section algorithm can be used to the synchronization of two above remained noncommutative Henon-Heiles
systems. In our treatment the canonically deformed model \cite{d1} plays the role of master system
\begin{eqnarray}
\dot{x}_1 &=& \left[{1}/{M({\theta})}\right]p_1 + S({\theta})x_2 + \nonumber\\
&~&~~~~~~~~~~~~~~~~~+\;\left[\left(x_1- {\theta}p_2\right)^2 - \left(x_2+ {\theta}p_1\right)^2\right]{\theta}\;,\label{canon1}\\[10pt]
\dot{x}_2 &=& \left[{1}/{M({\theta})}\right]p_2 - S({\theta})x_1 - 2(x_2+\theta p_1)\left(x_1- {\theta}p_2\right){\theta}\;,\\[10pt]
\dot{p}_1 &=& -{M({\theta})}\Omega^2({\theta})x_1 + S({\theta})p_2 -
2\left(x_2+ {\theta}p_1\right)(x_1-\theta p_2)\;,\\[10pt]
\dot{p}_2 &=& -{M({\theta})}\Omega^2({\theta})x_2 - S({\theta})p_1 + \;\nonumber\\
&~&~~~~~~~~~~~~~~~~~~~~~~~~-
\left(x_1- {\theta}p_2\right)^2 +
\left(x_2+ {\theta}p_1\right)^2\label{canon4}\;.
\end{eqnarray}
while its slave partner is given by Lie-algebraically noncommutative model \cite{d2}
\begin{eqnarray}
\dot{y}_1 &=& {1}/{M\left(\frac{t}{\kappa}\right)}\pi_1 + S\left(\frac{t}{\kappa}\right)y_2 +\nonumber\\
&~&~~~~~~~~~~~~~~~~~+\;
\left[\left(y_1- \frac{t}{\kappa}\pi_2\right)^2 - \left(y_2+ \frac{t}{\kappa}\pi_1\right)^2\right]\frac{t}{\kappa}+u_{y_1}\label{times1}\;,\\[5pt]
\dot{y}_2 &=& {1}/{M\left(\frac{t}{\kappa}\right)}\pi_2 - S\left(\frac{t}{\kappa}\right)y_1 - 2\left[y_2+ \frac{t}{\kappa}\pi_1\right]
\left[y_1- \frac{t}{\kappa}\pi_2\right]\frac{t}{\kappa} + u_{y_2}\label{times1a}\;,\\[5pt]
\dot{\pi}_1 &=& -{M\left(\frac{t}{\kappa}\right)}\Omega^2\left(\frac{t}{\kappa}\right)y_1 + S\left(\frac{t}{\kappa}\right)\pi_2 -
2\left[y_2+ \frac{t}{\kappa}\pi_1\right]\left[y_1- \frac{t}{\kappa}\pi_2\right]+ u_{\pi_1}\;,\\[5pt]
\dot{\pi}_2 &=& -{M\left(\frac{t}{\kappa}\right)}\Omega^2\left(\frac{t}{\kappa}\right)y_2 -
S\left(\frac{t}{\kappa}\right)\pi_1 +\nonumber\\
&~&~~~~~~~~~~~~~~~~~~~~~~~~~-\;
\left[y_1- \frac{t}{\kappa}\pi_2\right]^2 +
\left[y_2+ \frac{t}{\kappa}\pi_1\right]^2 + u_{\pi_2}\label{times4}\;,
\end{eqnarray}
with active controllers $u_{y_1}$, $u_{y_2}$, $u_{\pi_1}$ and $u_{\pi_2}$ respectively.

Using the above equations of motion one can check that
the dynamics of synchronization errors $e_{y_i} = y_i - x_i$ and $e_{\pi_i} = \pi_i - p_i$ is obtained
as\footnote{See also formula (\ref{eq4}).}
\begin{eqnarray}
\dot{e}_{y_1} &=&{1}/{M\left(\frac{t}{\kappa}\right)}\pi_1 + S\left(\frac{t}{\kappa}\right)y_2 +\nonumber\\
&~&~~~~~~~~~~~~~~~~~+\;
\left[\left(y_1- \frac{t}{\kappa}\pi_2\right)^2 - \left(y_2+ \frac{t}{\kappa}\pi_1\right)^2\right]\frac{t}{\kappa} + \;\label{times11}\\[10pt]
&-&\frac{1}{M({\theta})}p_1 - S({\theta})x_2 -\;\left[\left(x_1- {\theta}p_2\right)^2 + \left(x_2+ {\theta}p_1\right)^2\right]{\theta} + u_{y_1}
\nonumber\;\\[10pt]
&~~&~\cr
\dot{e}_{y_2} &=&  {1}/{M\left(\frac{t}{\kappa}\right)}\pi_2 - S\left(\frac{t}{\kappa}\right)y_1 - 2\left[y_2+ \frac{t}{\kappa}\pi_1\right]
\left[y_1- \frac{t}{\kappa}\pi_2\right]\frac{t}{\kappa}+\nonumber\;\\[5pt]
&~&~~~~~~~~~~~-\;\frac{1}{M({\theta})}p_2 + S({\theta})x_1 + 2(x_2+\theta p_1)\left(x_1- {\theta}p_2\right){\theta} + u_{y_2}\;,\\[10pt]
&~~&~\cr
\dot{e}_{\pi_1} &=& -{M\left(\frac{t}{\kappa}\right)}\Omega^2\left(\frac{t}{\kappa}\right)y_1 + S\left(\frac{t}{\kappa}\right)\pi_2 -
2\left[y_2+ \frac{t}{\kappa}\pi_1\right]\left[y_1- \frac{t}{\kappa}\pi_2\right]+ \nonumber\;\\[10pt]
&+&\;{M({\theta})}\Omega^2({\theta})x_1 - S({\theta})p_2 +
2\left(x_2+ {\theta}p_1\right)(x_1-\theta p_2) + u_{\pi_1}\;,\\[5pt]
&~~&~\cr
\dot{e}_{\pi_2} &=& -{M\left(\frac{t}{\kappa}\right)}\Omega^2\left(\frac{t}{\kappa}\right)y_2 -
S\left(\frac{t}{\kappa}\right)\pi_1 +\nonumber\\
&~&~~~~~~~~~~~~~~~~~~~~~~~~~~~~~-\;
\left[y_1- \frac{t}{\kappa}\pi_2\right]^2 +
\left[y_2+ \frac{t}{\kappa}\pi_1\right]^2 + \nonumber\;\\[5pt]
&+&\;{M({\theta})}\Omega^2({\theta})x_2 + S({\theta})p_1 + \left(x_1- {\theta}p_2\right)^2 -
\left(x_2+ {\theta}p_1\right)^2+ u_{\pi_2}\label{times44}\;.
\end{eqnarray}
Besides, if we define the positive Lyapunov function by\footnote{The matrix $P = 1$ in the formula (\ref{eq6}).}
\begin{eqnarray}
{V}(e) = \frac{1}{2}\left(e_{y_1}^2+e_{y_2}^2+e_{\pi_1}^2+e_{\pi_2}^2\right) \;, \label{eq11}
\end{eqnarray}
then for the following choice of control functions
\begin{eqnarray}
{u}_{y_1} &=& \left[{1}/{M({\theta})}\right]p_1 + S({\theta})x_2 +\left[\left(x_1- {\theta}p_2\right)^2 - \left(x_2+ {\theta}p_1\right)^2\right]{\theta}+\nonumber\\
&-&{1}/{M\left(\frac{t}{\kappa}\right)}\pi_1 - S\left(\frac{t}{\kappa}\right)y_2 + \label{controlers1}\\
&~&~~~~~~~~~~~~~~~~~~~~~~~~-\;
\left[\left(y_1- \frac{t}{\kappa}\pi_2\right)^2 + \left(y_2+ \frac{t}{\kappa}\pi_1\right)^2\right]\frac{t}{\kappa}
- e_{y_1} \;\nonumber\\[10pt]
&~~&~\cr
{u}_{y_2} &=& \left[{1}/{M({\theta})}\right]p_2 - S({\theta})x_1 - 2(x_2+\theta p_1)\left(x_1- {\theta}p_2\right){\theta} +\nonumber\\
&-&{1}/{M\left(\frac{t}{\kappa}\right)}\pi_2 + S\left(\frac{t}{\kappa}\right)y_1 + 2\left[y_2+ \frac{t}{\kappa}\pi_1\right]
\left[y_1- \frac{t}{\kappa}\pi_2\right]\frac{t}{\kappa} - e_{y_2}\;,\\[10pt]
&~~&~\cr
{u}_{\pi_1} &=&  -{M({\theta})}\Omega^2({\theta})x_1 + S({\theta})p_2 -
2\left(x_2+ {\theta}p_1\right)(x_1-\theta p_2) + \nonumber\\
&+&{M\left(\frac{t}{\kappa}\right)}\Omega^2\left(\frac{t}{\kappa}\right)y_1 - S\left(\frac{t}{\kappa}\right)\pi_2 +
2\left[y_2+ \frac{t}{\kappa}\pi_1\right]\left[y_1- \frac{t}{\kappa}\pi_2\right]
- e_{\pi_1}\;,\\[10pt]
&~~&~\cr
{u}_{\pi_2} &=&  -{M({\theta})}\Omega^2({\theta})x_2 - S({\theta})p_1 - \left(x_1- {\theta}p_2\right)^2 +
\left(x_2+ {\theta}p_1\right)^2+\nonumber\\
&+&{M\left(\frac{t}{\kappa}\right)}\Omega^2\left(\frac{t}{\kappa}\right)y_2 +
S\left(\frac{t}{\kappa}\right)\pi_1 +\label{controlers4}\\
&~&~~~~~~~~~~~~~~~~~~~~~~~~~~~~~~~~~~~~~+\;
\left[y_1- \frac{t}{\kappa}\pi_2\right]^2 +
\left[y_2+ \frac{t}{\kappa}\pi_1\right]^2
- e_{\pi_2}\nonumber \;,
\end{eqnarray}
we have\footnote{The matrix $Q = 1$ in the formula (\ref{eq7}).}
\begin{eqnarray}
\dot{V}(e) = -\left(e_{y_1}^2+e_{y_2}^2+e_{\pi_1}^2+e_{\pi_2}^2\right) \;. \label{eq13}
\end{eqnarray}
Such a result means (see general prescription) that the canonically (see (\ref{canon1})-(\ref{canon4})) and Lie-algebraically
(see (\ref{times1})-(\ref{times4})) Henon-Heiles systems are synchronized for all initial conditions with active controllers
(\ref{controlers1})-(\ref{controlers4}).

Let us now illustrate the above considerations by the proper numerical calculations.

First of all, we solve canonically deformed system (\ref{canon1})-(\ref{canon4}) with $\theta = 1$ as well as
we integrate the Lie-algebraically model (\ref{times1})-(\ref{times4}) for $\kappa=1$ and without active controllers
$u_{y_1}$, $u_{y_2}$, $u_{\pi_1}$ and $u_{\pi_2}$, for two different sets of initial conditions
\begin{eqnarray}
(x_{01}, x_{02};p_{01},p_{02}) = (0.01, -0.01; 0, 0)\;, \label{eq14}
\end{eqnarray}
and
\begin{eqnarray}
(y_{01}, y_{02};\pi_{01},\pi_{02}) = (0, 0; -0.02, 0.02)\;, \label{eq15}
\end{eqnarray}
respectively. The results are presented on {\bf Figure 1} - one can see that there exist
(in fact) the divergences between both phase space trajectories. Next, we find the solutions for the
master system (\ref{canon1})-(\ref{canon4}) (the $(x,p)$-trajectory) and for its slave partner (\ref{times1})-(\ref{times4}) with active controllers
(\ref{controlers1})-(\ref{controlers4}) (the $(y,\pi)$-trajectory) for initial data (\ref{eq14}) and (\ref{eq15}) respectively. Now, we see that the
corresponding phase space trajectories become synchronized - the vanishing in time error functions $e_{y_i}$ and $e_{\pi_i}$ are presented on {\bf Figure 2}. Additionally, we repeat the above numerical procedure for two another sets of initial data: $(x_0;p_0) = (0,0;0,0)$ and $(y_0;\pi_0)=(0.02,-0.02;0.01,-0.01)$; the obtained results are presented on {\bf Figures 3} and {\bf 4} respectively.

\section{Final remarks}

In this article we synchronize two noncommutative Henon-Heiles models with use of active control method.
Particularly, we find the proper active controllers (\ref{controlers1})-(\ref{controlers4}) as well as we perform numerical synchronization
of the systems for fixed values of deformation parameters $\theta$ and $\kappa$.

In our opinion the obtained result seems to be quite interesting due to the two reasons at least. Firstly, it finds the direct dynamical link between
two models defined on the completely different noncommutative space-times - the canonically twisted space and the  Lie-algebraically deformed space-time respectively. Such a connection suggests, that there may exist other, more fundamental (for example taken at the kinematical level) link between both, considered here systems. Secondly, it combines in quite matured way two disparate scientific fields, such as the elements of Quantum Group Theory with the techniques typical for the Classical Chaos domain.

Finally, it should be noted that the presented investigations can be extended in various ways.
For example, one may consider synchronization of the noncommutative Henon-Heiles models with use of other mentioned in
Introduction methods. Obviously, the works in this direction already started and are in progress.

\eject

\pagestyle{empty}
$~~~~~~~~~~~~~~~~~~$
\\
\\
\\
\\
\\
\\
\\
\begin{figure}[htp]
\includegraphics[width=\textwidth]{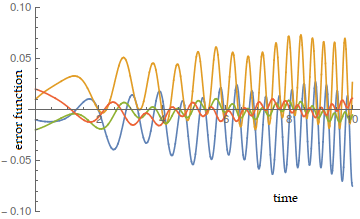}
\caption{The error functions $e_{y_i} = y_i - x_i$ and $e_{\pi_i} = \pi_i - p_i$ for canonically deformed Henon-Heiles system with
initial conditions (\ref{eq14}) (the $(x,p)$-trajectory), and for Lie-algebraically noncommutative Henon-Heiles model without correlation functions $u_{y_i}$, $u_{\pi_i}$
for the initial conditions (\ref{eq15}) (the $(y,\pi)$-trajectory). The blue line
corresponds to the $e_{y_1}$-error function, the orange one - to $e_{y_2}$, the green one - to $e_{\pi_1}$ and the red one - to $e_{\pi_2}$ respectively.}\label{grysunek1}
\end{figure}
\begin{figure}[htp]
\includegraphics[width=\textwidth]{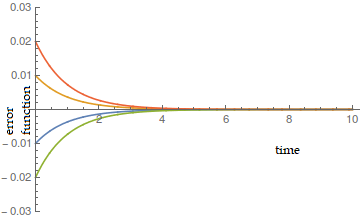}
\caption{The error functions $e_{y_i} = y_i - x_i$ and $e_{\pi_i} = \pi_i - p_i$ for canonically deformed Henon-Heiles model defined by the master system (\ref{canon1})-(\ref{canon4}) with the
initial conditions (\ref{eq14}) (the $(x,p)$-trajectory), and for the slave Lie-algebraically noncommutative Henon-Heiles system
(\ref{times1})-(\ref{times4}) with the initial conditions (\ref{eq15}) (the $(y,\pi)$-trajectory).
The blue line
corresponds to the $e_{y_1}$-error function, the orange one - to $e_{y_2}$, the green one - to $e_{\pi_1}$ and the red one - to $e_{\pi_2}$ respectively.}\label{grysunek2}
\end{figure}
\begin{figure}[htp]
\includegraphics[width=\textwidth]{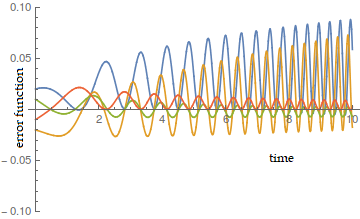}
\caption{The error functions $e_{y_i} = y_i - x_i$ and $e_{\pi_i} = \pi_i - p_i$ for canonically deformed Henon-Heiles model
 with the
initial conditions $(x_0;p_0) = (0,0;0,0)$ (the $(x,p)$-trajectory), and for Lie-algebraically noncommutative Henon-Heiles  model without correlation functions $u_{y_i}$, $u_{\pi_i}$ for
the initial conditions $(y_0,\pi_0)=(0.02,-0.02;0.01,-0.01)$ (the $(y,\pi)$-trajectory). The blue line
corresponds to the $e_{y_1}$-error function, the orange one - to $e_{y_2}$, the green one - to $e_{\pi_1}$ and the red one - to $e_{\pi_2}$ respectively.}\label{grysunek3}
\end{figure}
\begin{figure}[htp]
\includegraphics[width=\textwidth]{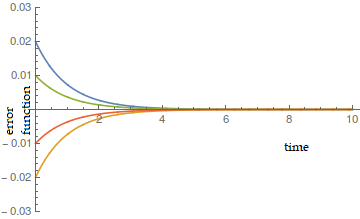}
\caption{The error functions $e_{y_i} = y_i - x_i$ and $e_{\pi_i} = \pi_i - p_i$ for canonically deformed Henon-Heiles model defined by the master system (\ref{canon1})-(\ref{canon4}) with the
initial conditions $(x_0;p_0) = (0,0;0,0)$ (the $(x,p)$-trajectory), and for the slave Lie-algebraically noncommutative Henon-Heiles system
(\ref{times1})-(\ref{times4}) with the initial conditions $(y_0,\pi_0)=(0.02,-0.02;0.01,-0.01)$
(the $(y,\pi)$-trajectory). The blue line
corresponds to the $e_{y_1}$-error function, the orange one - to $e_{y_2}$, the green one - to $e_{\pi_1}$ and the red one - to $e_{\pi_2}$ respectively.}
\label{grysunek4}
\end{figure}

\end{document}